\documentclass{article}

\PassOptionsToPackage{square,numbers, compress}{natbib}
\usepackage[final]{neurips_2020}

\usepackage[dvipsnames]{xcolor}
\usepackage[utf8]{inputenc} % allow utf-8 input
\usepackage[T1]{fontenc}    % use 8-bit T1 fonts
\usepackage{hyperref}       % hyperlinks
\usepackage{url}            % simple URL typesetting
\usepackage{booktabs}       % professional-quality tables
\usepackage{amsfonts}       % blackboard math symbols
\usepackage{nicefrac}       % compact symbols for 1/2, etc.
\usepackage{microtype}      % microtypography
\usepackage{bm}
\usepackage{multirow}
\usepackage{setspace}
\usepackage[square,numbers]{natbib}

\usepackage{titlesec}
\titlespacing*{\section}{0.0cm}{0.1cm}{0.1cm}
 
\usepackage[pdftex]{graphicx}

\title{Steerable discovery of neural audio effects}

\author{% 
  Christian J. Steinmetz \\
  \texttt{c.j.steinmetz@qmul.ac.uk} \\
  \And 
  Joshua D. Reiss \\
  \texttt{joshua.reiss@qmul.ac.uk} \\
  \And \vspace{-0.6cm} \\ 
  Centre for Digital Music, Queen Mary University of London, London, UK    
}

\begin{document}
 
\maketitle
\vspace{-0.5cm}
\begin{abstract}
\vspace{-0.1cm}
Applications of deep learning for audio effects often focus on modeling analog effects or learning to control effects to emulate a trained audio engineer. 
However, deep learning approaches also have the potential to expand creativity through neural audio effects that enable new sound transformations. 
While recent work demonstrated that neural networks with random weights produce compelling audio effects, control of these effects is limited and unintuitive.
To address this, we introduce a method for the steerable discovery of neural audio effects.
This method enables the design of effects using example recordings provided by the user. 
We demonstrate how this method produces an effect similar to the target effect, along with interesting inaccuracies, while also providing perceptually relevant controls.
\vspace{-0.0cm}
\end{abstract}

\section{Introduction}

% first paragraph sets the scene for what we want to do, high level and history.

%While the advancement of music technology continually introduces new tools and techniques to address challenges in music production, the intentional misuse of this technology often gives birth to new styles and genres. 
%For example 
%From the early blues guitarists who experimented with pushing amplifiers beyond their limits to achieve distortion~\cite{shepherd2003distortion}, to the DJ's use of turntables as an instrument, introducing ``scratching'' technique~\cite{hansen2002basics}, and the adoption of Auto-Tune as a vocal augmentation~\cite{diaz2009autotune}, exploitation of music technology in ways the original designers never imagined is an important pattern in the advancement of music~\cite{prior2009software}.
%While applications of machine learning in music production continue to grow, with promise of automatic systems for music composition~\cite{}, synthesis~\cite{dhariwal2020jukebox, engel2020ddsp}, and audio engineering~\cite{martinez2021deep, steinmetz2020mixing, choi2021amss}, it is not yet clear how these system may be exploited for creative effect.

% second paragraph goes into what we have done so far, we can note things like timbre transfer (synthesis), but work on audio effects is so far much more limited. Here we can setup our formulation of ronn, which is unconditional, and establish the idea of the weight space and architecture space 

Audio effects are specialized signal processing tools used in music production for shaping the loudness, timbre, pitch, spatialization, or rhythm of sound~\cite{wilmering2020history}.
There has been growing interest in deep learning for emulating analog audio effects~\cite{damskagg2019distortion, martinez2019nonlinear, wright2020real, signaltrain, steinmetz2021efficient}, as well as methods for automatic control of audio effects to simplify music production~\cite{sheng2019feature, mimilakis2020one, ramirez2021differentiable, steinmetz2020mixing}.
In addition, related work in neural audio synthesis has investigated how deep learning may enable the synthesis of new sounds~\cite{engel2017neural, hantrakul2019fast, engel2020ddsp, hayes2021neural}. 
The expressive modeling capability of neural networks may enable audio effects that have not yet been realized by traditional signal processing approaches. 
However, deep learning methods for discovering new kinds of neural audio effects that expand or augment creative options for musicians remain limited.

Recent work demonstrated that neural networks with randomly initialized weights produce a range of compelling audio effects~\cite{steinmetz2020overdrive}.
Furthermore, architectural details have an impact on the resulting effect, with deeper models leading to reverb and delay-like effects, and shallower networks leading to roomy distortion-like effects. 
However, this approach for generating neural audio effects is limited in two main ways. 
First, the resulting effects are largely dependent on the distributions from which the random weights are sampled.
As a result, potentially interesting points in the weight space of these networks may never be reached. 
Second, while there is a connection between the architecture and the resulting effect, adjusting these attributes towards a sonic goal remains a challenging trial and error process.
%Additionally, the stochastic element of this process means that many random iterations are often required to achieve an interesting effect.
These challenges motivated our proposed method for a steerable generation process that allows users to direct the design of new effects according to their own aesthetic goals. 

Most deep learning applications involve a clearly defined task, for example, emulating an audio effect as accurately as possible.
However, the task of discovering new effects is clearly open-ended.
In this case, we desire an approach that balances producing an ``interesting'' audio effect while ensuring this effect provides some level of intuitive  (often perceptual) user control. 
We propose a simple steering method that uses a single, short input recording, along with a version of this recording processed by an audio effect similar to the one we would like to construct. 
We then train a conditional temporal convolutional network (TCN)~\cite{ steinmetz2021efficient} on this single input/output pair while holding the conditioning constant.
After only a few minutes this process produces a rough emulation of the target effect with some interesting inaccuracies characteristic of convolutional neural networks. 
%When applying this effect to other signals, further interesting inaccuracies often arise.
Surprisingly, we find that adjusting the conditioning signal to values other than those seen during the steering process results in controls that tends to be correlated with perceptual attributes of the neural audio effect. 

% third paragraph we give the intro to our new idea, which is a method for steering this process.

\begin{figure}[]
    \centering
    \vspace{-0.8cm}
    \includegraphics[trim={0.5cm 0 0.05cm 0},clip, width=\linewidth]{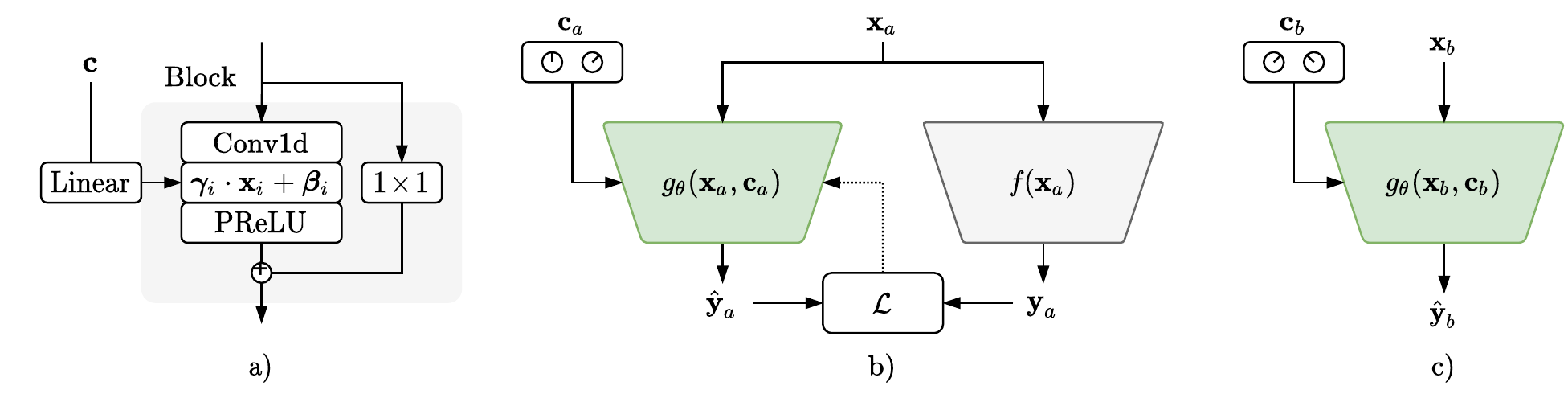}
    \vspace{-0.5cm}
    \caption{a) TCN block with 1D convolution, conditional affine transformation (FiLM), followed by a PReLU nonlinearity.
    b) Steering process where $g_\theta(\textbf{x}_a,\textbf{c}_a)$, a conditional TCN, is trained to emulate $f(\textbf{x}_a)$, an existing audio effect, using a single input/output pair of recordings $\textbf{x}_a, \textbf{y}_a$.
    c) Generation process where $\textbf{x}_b$, a new signal, is processed with the TCN and differing conditioning parameters $\textbf{c}_b$.}
    \label{fig:block_diagram}
    \vspace{-0.5cm}
\end{figure}

\section{Method}

% start this section by diving into how we adapt the formulation of ronn to be conditional.

We begin by considering $g_\theta(\textbf{x}_a, \textbf{c}_a)$, a conditional TCN with weights $\theta \in \mathbb{R}^{P}$. This network processes an audio signal $\textbf{x}_a \in \mathbb{R}^{C \times N}$ with $C$ channels and $N$ samples according to the conditioning signal $\textbf{c}_a \in \mathbb{R}^{D}$ to produce a transformed signal $\hat{\textbf{y}}_a \in \mathbb{R}^{C \times N}$. 
Our proposed steering process, shown in Figure~\ref{fig:block_diagram}b, requires an input signal $\textbf{x}_a$ along with a version of this signal $\textbf{y}_a$ that has been processed by another effect $f(\textbf{x}_a)$, for example an existing reverb, dynamic range compressor, or other effect. 
We set the conditioning parameters $\textbf{c}_a = \mathbf{0}$ and hold them constant during the steering process. 
Then we train the model using gradient descent, iteratively updating the weights $\theta := \theta - \eta \nabla_{\theta} \mathcal{L} \big( g_{\theta}(\textbf{x}_a, \textbf{c}_a), \textbf{y}_a \big) $ with learning rate $\eta$, so the output $\hat{\textbf{y}}_a$ is close to the processed signal $\textbf{y}_a$ according to some loss function $\mathcal{L}$.
After steering, we can use this network to process other audio signals, as shown in Figure~\ref{fig:block_diagram}c. 
Additionally, we can adjust the conditioning signal to another value $\textbf{c}_b$, which will change the resulting effect by modulating the intermediate features in a different manner.

We utilize the formulation of the TCN presented in~\cite{steinmetz2021efficient}, which employs residual blocks as shown in Figure~\ref{fig:block_diagram}a. 
In order to integrate conditioning, we use feature-wise linear modulation (FiLM)~\cite{perez2018film}.
To implement this, intermediate features at the $i$-th layer $\textbf{x}_i$, with shape $B \times C \times N$ (batch, channels, samples), are modulated by a set of scaling $\bm{\gamma}_i$ and shifting $\bm{\beta}_i$ parameters.
These parameters are derived from the global conditioning signal $\textbf{c}$ which is projected from $\mathbb{R}^{D}$ to $\mathbb{R}^{2 \cdot C}$ at each intermediate layer.
Finally, a PReLU~\cite{he2016deep} nonlinearity is applied. 
During optimization, we use Adam along with a formulation of the multiresolution STFT~\cite{yamamoto2020parallel} for $\mathcal{L}$ based on the implementation in \texttt{auraloss}~\cite{steinmetz2020auraloss}. 
%with window sizes $ w_n \in { 512, 1024, 2048} $.

\section{Results}
% This will contain experiments, results and conclusions
%In essence, this is a special case of audio effect modeling where only a single example is used for training. 
%In most cases, this will not create an emulation that is true to the original effect.
%However, we find that the resulting neural audio effect often contains similar characteristics.
%This steering process enables us to locate ``random'' points within the weight space $\mathbb{R}^{P}$ of network, yet 

We applied this steering process with examples from a compressor, analog delay, guitar amplifier, and reverberation effect. 
We used $\textbf{c} \in \mathbb{R}^{2}$ to provide 2 control parameters and trained for 2500 steps using recordings at $44.1$\,kHz.
We found it best to match the receptive field of the network to the expected memory of the effect.
For example, we used a receptive field of $\approx 300$\,ms for the compressor and $\approx 2000$\,ms for reverberation. 
This process resulted in an effect with similar characteristics to the original effects, however with some inaccuracies such as interesting delay-like patterns and distortions.
Furthermore, we found that varying the control parameters tends to change relevant aspects of the effect, such as the amount of compression or the decay time.
We provide a Colab notebook\footnote{\url{https://csteinmetz1.github.io/steerable-nafx}\vspace{-0.4cm}} for these experiments along with listening examples that demonstrate the potential range of these effects. 

\begin{figure}[h]
    \vspace{-0.5cm}
    \centering
    \begin{minipage}{0.5\textwidth}
        \centering
        \includegraphics[trim={0cm 0cm 0cm 1cm},clip, width=0.9\textwidth]{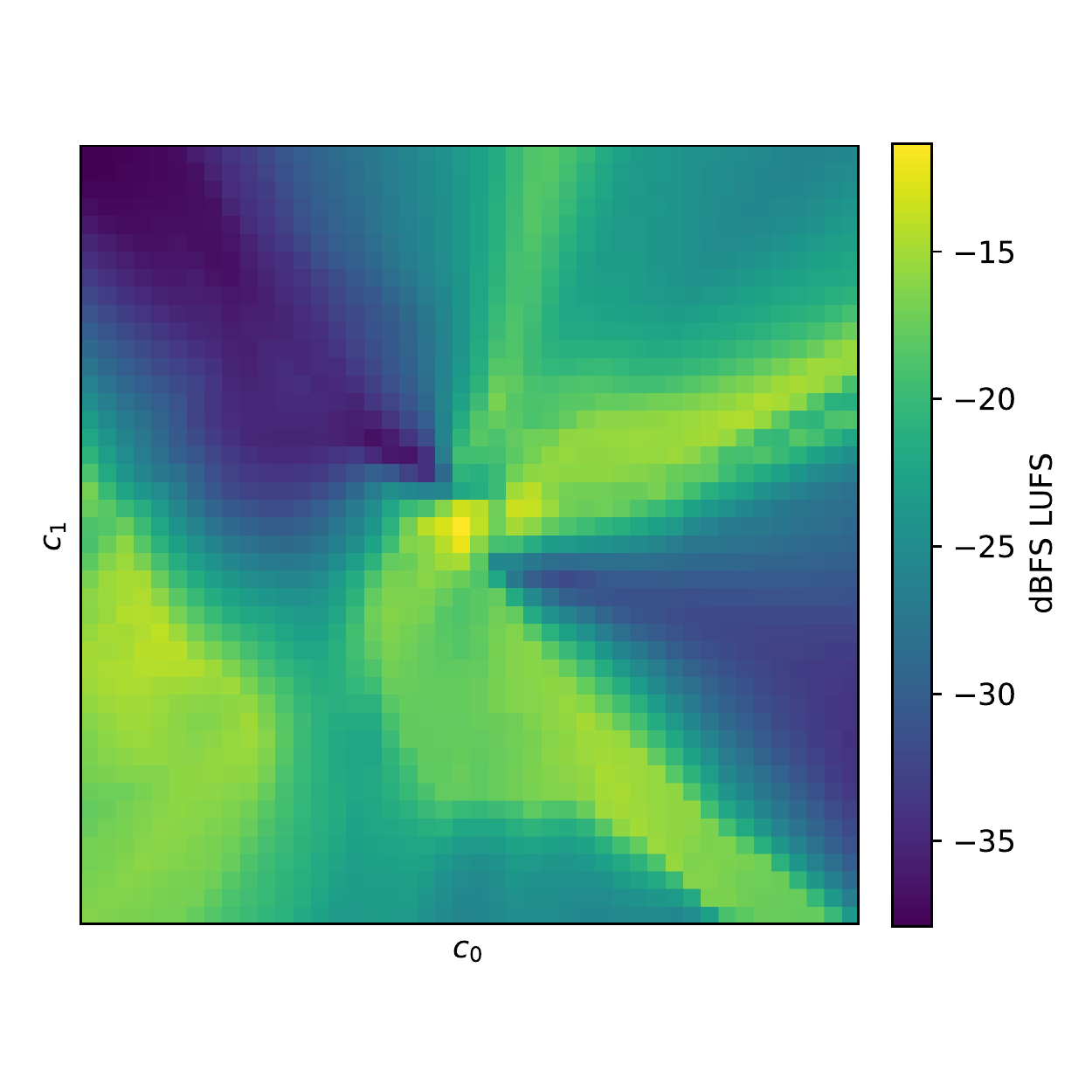} % first figure itself
        \vspace{-0.4cm} \\
        \hspace{-1cm}\centering{a) Dynamic range compressor}
    \end{minipage}\hfill
    \begin{minipage}{0.5\textwidth}
        \centering
        \includegraphics[trim={0cm 0cm 0cm 1cm},clip,width=0.9\textwidth]{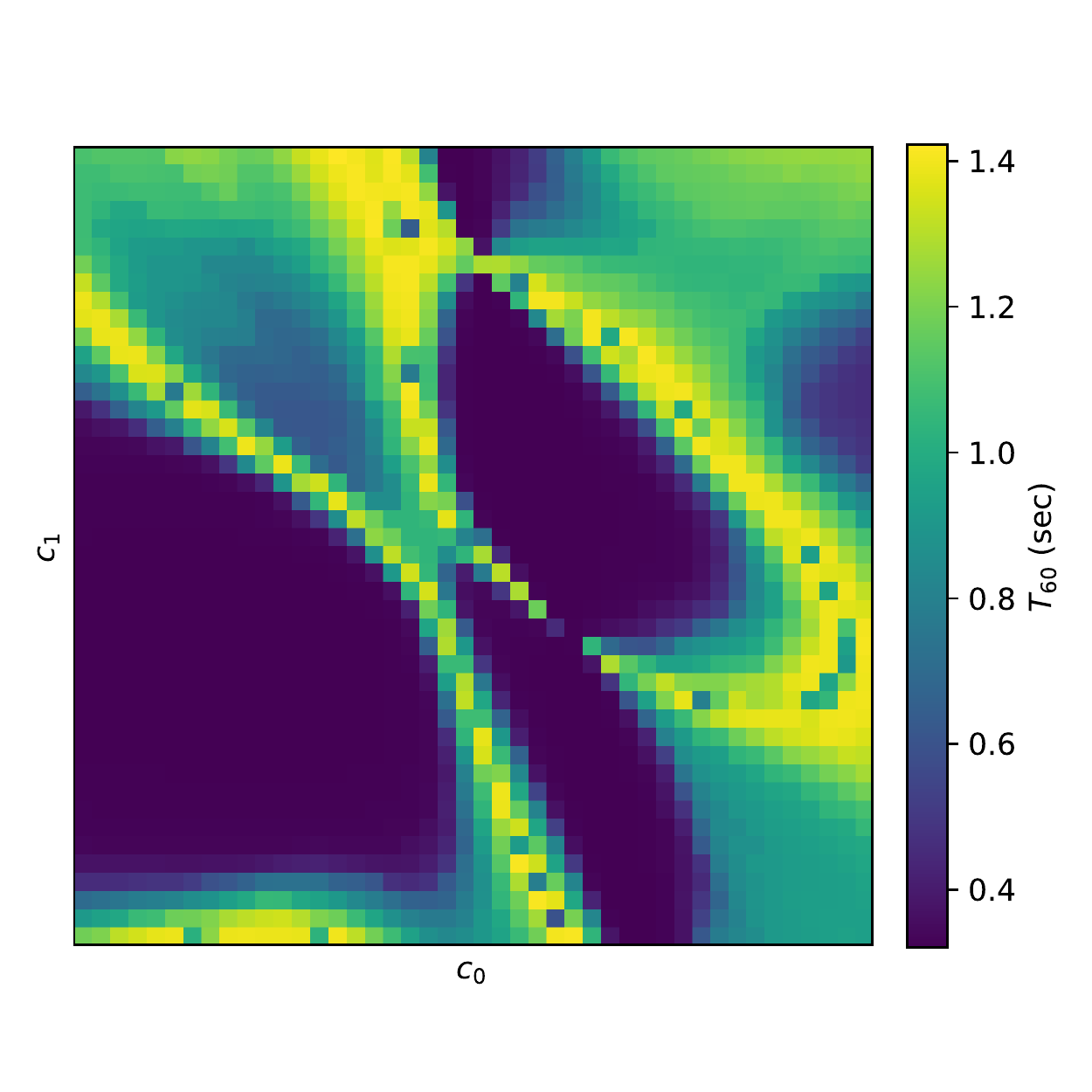} % second figure itself
        \vspace{-0.4cm} \\
        \hspace{-1cm}\centering{b) Artificial reverberation}
    \end{minipage}
    \caption{Parameter space $\textbf{c} \in \mathbb{R}^{2}$ from $-5$ to $5$ with relation to a) loudness dB LUFS for a model steered with a signal from a dynamic range compressor, and b) $T_{60}$ for a model steered with a signal from an artificial reverberation effect, both of which demonstrate clear structure.}
    \label{fig:grids}
\end{figure}

\begin{figure}[h]
    \vspace{-0.2cm}
    \centering
   \includegraphics[width=0.8\linewidth]{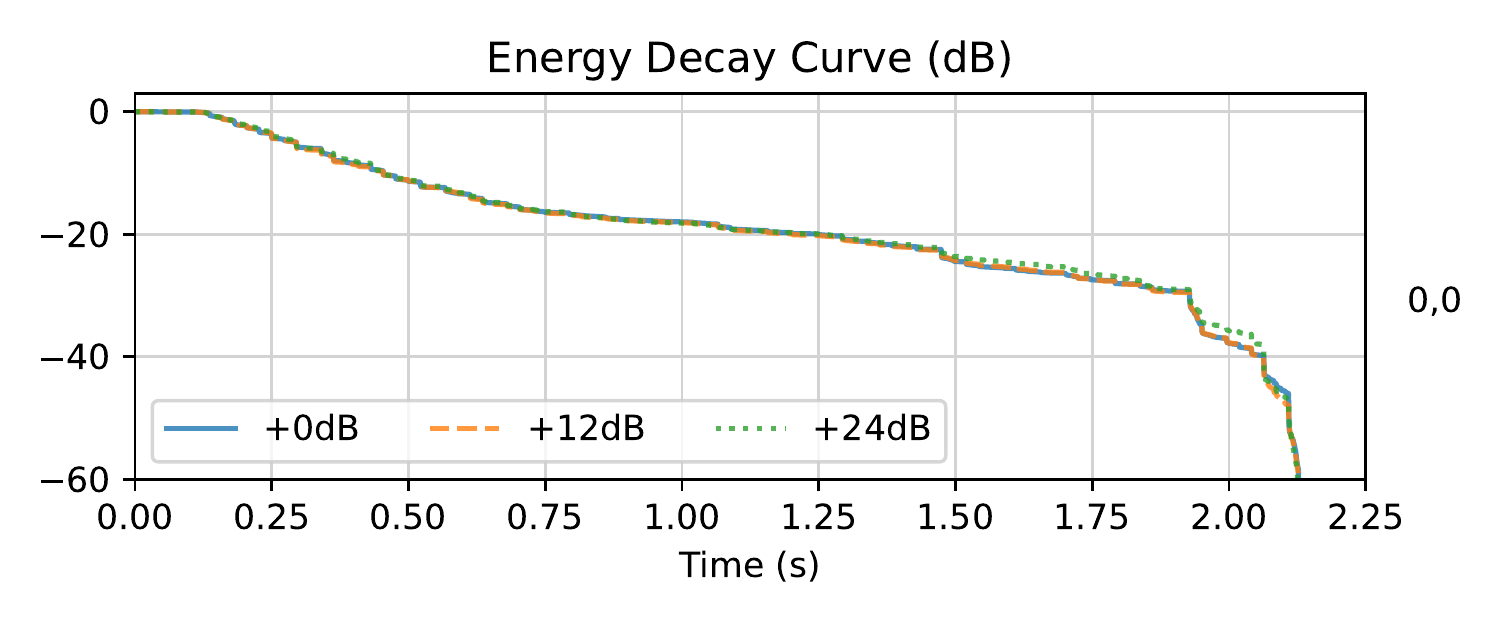}
    \vspace{-0.4cm}
    \caption{Multiple energy decay curves with an impulse at varying input levels as input to the model. Even though the TCN is nonlinear, the decay appears consistent as expected from a reverb-like effect.}
    \label{fig:decay}
    \vspace{-0.3cm}
\end{figure}

To further investigate the  perceptual relevance of the control parameters, we evaluate the model steered with the compressor signal by processing a piano recording, varying the conditioning $\textbf{c}$ over a $2\textrm{D}$ grid from $-5$ to $5$. 
We create a heatmap of the loudness in dB LUFS~\cite{bs1770}, as shown in Figure~\ref{fig:grids}a, which demonstrates clear structure in the control parameter space. 
We perform a similar experiment for the reverberation effect, but instead use an impulse as input, measuring the $T_{60}$, shown in Figure~\ref{fig:grids}b. 
Even though the TCN is nonlinear, we found the $T_{60}$ provided a sense of the reverberation length, with the input level having a limited effect on the decay as shown in Figure~\ref{fig:decay}.

These experiments demonstrate only a few applications of this approach.
Future work may involve using steering signals originating from other audio effects, as well as unconventional sources, such as learning a mapping from one sound to another.
The eventual goal would be to drive the steering process using only one signal, enabling the use of arbitrary recordings without the original clean signal. 
Other directions include experimentation with other architectures, as well as approaches for quantifying the ``interestingness'' of neural audio effects for use as an objective function. 

\newpage
\section*{Acknowledgement}

This work is supported by the EPSRC UKRI Centre for Doctoral Training in Artificial Intelligence and Music (EP/S022694/1).

%\section*{Broader Impact}

\bibliographystyle{ieeetr}
\bibliography{references}
  
\newpage 
\vspace{-1cm}
\section*{Supplementary materials}

\begin{table}[h]
    \vspace{0.6cm}
    \centering
    \begin{tabular}{l c c}
        \toprule
         & \multicolumn{2}{c}{\textbf{Models}} \\ \cmidrule(lr){2-3}
       \textbf{Parameter} & Compressor & Reverberation  \\ \midrule
        Layers & 4 & 5\\
        Channels & 32 & 32\\
        Kernel size & 9 & 9\\ 
        Dilation growth & 10 & 10 \\ 
        Learnable parameters & 21,803 & 32,268 \\ 
        \midrule 
        Receptive field (samples) & 8889 & 88889 \\ 
      \multicolumn{1}{r}{(ms)} & 201.6 & 2015.6\\ 
       \midrule 
        Learning rate & \multicolumn{2}{c}{1e-3, 1e-4 (80\%), 1e-5 (95\%)}\\
        %\multicolumn{1}{r}{Step 2000 (80\%)} & \multicolumn{2}{c}{1e-4}\\
        %\multicolumn{1}{r}{Step 2375 (95\%)} & \multicolumn{2}{c}{1e-5}\\
        Optimizer & \multicolumn{2}{c}{Adam}\\
        Iterations & \multicolumn{2}{c}{2500}\\
        MR-STFT sizes & \multicolumn{2}{c}{\{32, 128, 512, 2048\}} \\
       \midrule
        Train time & 4 min 26 sec & 7 min 4 sec\\ 
        MR-STFT Error & 0.4945 & 0.9453 \\ 
         \bottomrule
    \end{tabular}
    \vspace{0.3cm}
    \caption{Model hyperparameters}
    \label{tab:hyperparam}
    \vspace{-1.8cm}
\end{table}

\end{document}